# AUTOMATED GENERATION OF COMPUTER GRADED UNIT TESTING-BASED PROGRAMMING ASSESSMENTS FOR EDUCATION

Sébastien Combéfis[1, 2] and Guillaume de Moffarts[2]

[1]ECAM Brussels Engineering School, Brussels, Belgium
[2]Computer Science and IT in Education ASBL, Louvain-la-Neuve, Belgium

## ABSTRACT

*Automatic assessment of code, in particular to support education, is an important feature included in several Learning Management Systems (LMS), at least to some extent. Several kinds of assessments can be designed, such as exercises asking to "fill the following code", "write a function that", or "correct the bug in the following program", for example. One difficulty for instructors is to create such programming exercises, in particular when they are somewhat complex. Indeed, instructors need to write the statement of the exercise, think about the solution and provide all the additional information necessary to the platform to grade the assessment. Another difficulty occurs when instructors want to use their exercises on another LMS platform. Since there is no standard way to define and describe a coding exercise yet, instructors have to re-encode their exercises into the other LMS. This paper presents a tool that can automatically generate programming exercises, from one single and unique description, and that can be solved in several programming languages. The generated exercises can be automatically graded by the same platform, providing intelligent feedback to its users to support their learning. This paper focuses on and details unit testing-based exercises and provides insights into new kinds of exercises that could be generated by the platform in the future, with some additional developments.*

## 1. INTRODUCTION

Being able to automatically grade code produced by learners, and in particular students in schools and universities, is a very demanded feature for Learning Management Platforms (LMS) [1]. In particular, professors in charge of programming courses need to assess the programming skills of their students. It is of course also the case for other courses that may require some programming, such as data mining or natural language processing courses, for example. The main issue is that this assessment cannot be done manually, especially if there are a large number of students [2-3]. Another situation, where automatic code assessment is mandatory, is Massive Open Online Courses (MOOCs), for which students are spread all over the world and are even more numerous [4-5]. Of course, the automatic grading of code must be more advanced than just assessing whether the code compiles and produces the correct result for some test cases. It should provide useful feedback to the learners. It is even more important when the number of students is large or in the case of MOOCs for which learners have a more limited access to the professors to get more individualised feedbacks.

This paper proposes a tool to generate coding exercises that can be solved in several programming languages and that can be automatically graded. Exercises are generated from a single language-agnostic configuration file. The same configuration can therefore be used to generate several instances of the same exercise for different programming languages. Feedbacks generated by the tool, and provided to the learners, are designed to help the learners to identify and understand their faults. They are also more suited for education and designed to support their learning. The first prototype of the tool [6] has been used to support a university course, at the Université catholique de Louvain (UCLouvain), introducing students to programming concepts and paradigms, as well as for a MOOC on the same topic [4]. The more recently rewritten version adds generic unit testing-based exercises [7]. It has been used for a second bachelor course about Python programming at the ECAM Brussels Engineering School, a higher education institution for future engineers. Finally, the last version of the tool, presented in this paper, supports automatic generation of unit testing-based exercises. It is currently tested with EDITx, a private company that organises IT challenges targeted to IT students and IT professionals, all around Europe.

## 1.1. Motivation

A lot of tools that can automatically grade codes do exist. They can generally be split in three categories: (a) code grading for programming competitions (online or onsite), (b) code evaluation for test-driven development and (c) code grading for education. For competitions, it is important to be able to guarantee the same execution environment and conditions for all the code evaluations. The main reason being that code evaluations are used to establish the ranking and to offer prizes to the participants. For example, it should be possible to impose time and memory limits that cannot be exceeded during the execution of participants' code submissions. Those graders must also be very robust to hold on during the whole competition, and must guarantee code and grading traceability in case of complaints [8]. For development, programs are typically tested to check whether their code is functionally correct regarding the executed test cases, following the Test-Driven Development (TDD) approach. For such assessments, time and memory constraints are less useful, but defining and controlling the test environment is also important. It should also be important to test the same code under different situations, for example to evaluate some fault tolerance levels. Finally, when it comes to assess code for educational purposes, several additional requirements arise. First, the feedback provided to learners must support their learning and cannot be limited to the classical "pass/fail" verdict of standard graders. The feedbacks must help learners to understand their faults and to make progress. Then, graders for education must support a larger number of different execution environments and programming languages than competition or development graders, that are often more specific. Finally, learner's code must be executed in a safe environment, for example isolated in sandboxes, because learners may produce wrong or dangerous code, whether it is voluntary or not.

All these observations led to the development of Pythia, a platform that combines requirements from the three categories of graders presented above. This platform has been designed to support education and, in particular, the teaching and learning of programming [6-7]. The main motivation that gave birth to the Pythia platform is to propose a tool on which several kinds of programming exercises can be automatically graded. It must also be flexible enough so that codes produced by the learners can be thoroughly analysed with existing tools. Therefore, the Pythia platform can support various assessments based on several criterions (functional correctness, code quality, execution performance, memory consumption, etc.). Finally, the platform should allow instructors to easily produce exercises following existing templates, or to build their own exercises specifically tailored for their students. From the competition graders, Pythia took two ideas: isolated sandboxes to safely execute code and possibility to impose constraints (such as time and memory limits). From the "TDD graders", Pythia took the idea of the systematic way to test codes against test suites. Finally, from education graders, Pythia took the idea of working on tailored "intelligent" feedbacks that support learning.

## 1.2. Related Work

As detailed above, many code graders have been developed, but most of them are either competition graders or specific ones only being able to handle certain kinds of exercises [9]. Several reviews have been conducted and interested reader can refer to them [10-13]. Among those graders, some follows the "TDD grader" philosophy and are based on tests [13]. When graders are to be used for educational purposes, reviews agree that feedback is important and that good feedback helps the learners and support their learning [14-15]. Finally, concerning automatic generation of programming exercises, only some solutions have been developed, but it is important in particular in the case of large classes to be able to easily diversify the number of available exercises [16-17].

The remainder of the paper is structured as follows. Section 2 presents a global overview of the architecture of Pythia. Then, Section 3 presents how to define an exercise and how it will be generated. Finally, Section 4 concludes the paper with some discussions and future works.

## 2. PYTHIA PLATFORM ARCHITECTURE

Pythia is a distributed application with several components. It has mainly been developed with the Go programming language. It uses UML virtual machines to execute code in a safe and controlled environment. The details of the architecture of the Pythia platform not being the purpose of this paper, the interested reader can refer to [6], or can directly delve into its code available here: https://github.com/pythia-project, to get a better understanding of it. Figure 1 shows a global overview of the architecture of the Pythia platform. The client interacts with the platform through an API server. This latter is connected to the Pythia backend, which manages the code execution within virtual machines (VM). Tasks to be executed and environments in which tasks can be executed are available to the backend and API server. They are in fact SquashFS read-only file systems stored on disk as files (TaskDB and EnvDB).

Two scenario examples are illustrated on Figure 1:

1. An instructor can call a specific route on the API server to create a new task. The task generator component will create it and store it in the TaskDB.

2. A learner can call a specific route on the API server to execute a task. The submission grader component will execute the task with the submission of the learner on the backend and return the generated feedback to the learner.

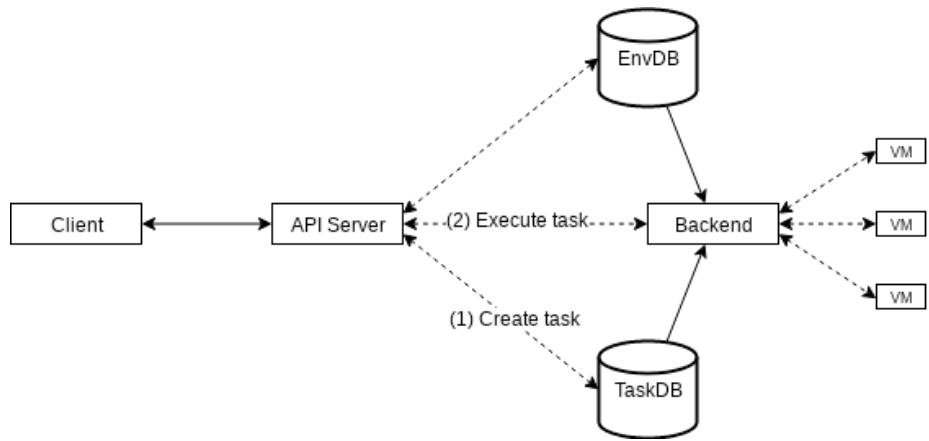

Figure 1. The Pythia platform is a distributed application with a backend managing VMs. It can be accessed through an API server that also manages tasks and execution environments.

## 2.1. Submission Grader

The submission grader is the component in charge of gathering the submission of the learner for a specific task and to evaluate it. The POST /api/execute route of the API server takes two parameters: the unique identifier of the task to execute and an input, which is a string containing the submission of the learner. The API server responds with three elements: the unique identifier of the task that has been executed, the status of the execution by the Pythia backend (success, timeout, overflow, etc.) and the output produced by the execution of the task (which contains among others the feedback). Depending on the kind of exercise, the input provided to the Pythia backend and the output produced by the execution of the task can be structured following a specified format. The Pythia platform does not impose anything on input and output. They just have to be strings, with a limitation on the number of characters for the output.

For example, Figure 2 shows the input and the output produced by the execution of a unit testing-based exercise where the student has to write the body of a function that computes the subtraction of its two arguments. The input should be a JSON object with two keys, one with a unique submission ID and one with the set of pieces of submitted code. For this particular exercise, there was only one field to fill out, named f1. The produced output contains the unique submission ID, the status of the execution of the tests (success, failed), and some feedback information. In this case, the feedback contains four elements:

- a score: 0.14285715,
- some statistics about the tests: 2 succeeded tests on a test suite with 14 tests,
- an example of inputs for which a test failed: for input (10,5), the expected answer is 5 (that is, $10 - 5$) and the answer computed by the learner's code is 10, and finally
- a message to help the learner find his/her fault: "Have you subtracted the 2nd parameter?".

The score and the statistics help the learner to evaluate how far from the completion of the exercise he/she is. The goal is to reach a score of 1, that is, to succeed all the tests from the test suite. The learner can also evaluate his/her own progress between submissions for the same exercise thanks to those statistics. Thanks to the example of inputs for which a test failed, the learner can trace his/her code execution to understand why it produced a wrong result. The learner can also check his/her corrected code before submitted it again, thanks to the provided expected answer. Finally, the message associated to the example of input should help the learner to find his/her fault. Since this message is more intuitive and related to the statement of the exercise to solve, it should encourage the learner to think about his/her solution, and not to try to change the code just to pass the failed test.

The handling of this specific input and the generation of this specific output are managed by code embedded in this particular task. In the Pythia platform, a task is in fact just a bunch of code that is executed in a safe environment, namely the UML virtual machine, taking a string as input and producing a string as output. An instructor can therefore create any kind of exercise, as long as he/she is able to write a code to parse the provided input, to evaluate the learner submission and to produce an output. He/she also has to define precise specifications for the input provided to his/her task and the generated output. Since the execution takes place inside a Linux virtual machine, the instructor can use any existing tool running on Linux to write a task. The only flip side of such flexibility is that creating a task can be very time-consuming and limited to only some instructors that have high programming skills and that understand the internal working of Pythia environments and tasks.

```
===INPUT EXAMPLE===
{
    "tid": "sub",
    "input": "{\"tid\": \"s001\", \"fields\": {\"f1\": \"return a\"}}"
}

===OUTPUT EXAMPLE===
{
    "tid": "s001",
    "status": "failed",
    "feedback": {
        "example": {
            "input": "(10,5)",
            "expected": "5",
            "actual": "10"
        },
        "message": "Have you subtracted the 2nd parameter?",
        "stats": {
            "succeeded": 2,
            "total": 14
        },
        "score": 0.14285715
    }
}
```

Figure 2. The execution of a unit testing-based task requires specific input information with the pieces of submitted code and produces a specific output with "intelligent" feedback information.

## 2.2. Task Generator

The task generator is a component in charge of automating the creation of tasks based on predefined templates. It can be used to ease the creation of exercises on the Pythia platform for instructors. For that, task templates must be defined, that is, a highly configurable generic program with placeholders must be designed as a task. Unit testing-based tasks [7] are structured following four processes as shown on Figure 3. The execution goes as follows:

1. The input of the learner is pre-processed, and used to fill a template code to produce the student code. This first step also initialises several files and directories. For example, it saves the task ID (tid) in a text file so that it can be used at the end of the task execution to generate the output of the task.

2. A test suite is then automatically generated based on the test configuration of the task contained in the test.json file. This file contains a set of predefined tests and configuration information to generate random tests. The test suite is stored with the CSV format in the data.csv file.

3.    The student code is then executed for each test of the test suite and the results of each execution are stored in the data.res text file. Each line of this file contains the verdict of the execution (checked, exception, etc.) with an associated value (the produced result, the description of the exception, etc.). Student code is executed in an unprivileged mode inside the virtual machine, so that it cannot access the correct solution or view some configuration files, for example.

4.    Finally, the correct solution stored in the solution.json file is fed in the template code to produce the teacher code, which is executed to produce the correct solutions for the generated test suite. Solutions are stored in the solution.res text file, each line containing the correct answer for each test. Then, the feedback is generated, comparing the correct answers with the ones produced by the learner. The test.json file is again used, to get information about the predefined tests and customised feedback messages.

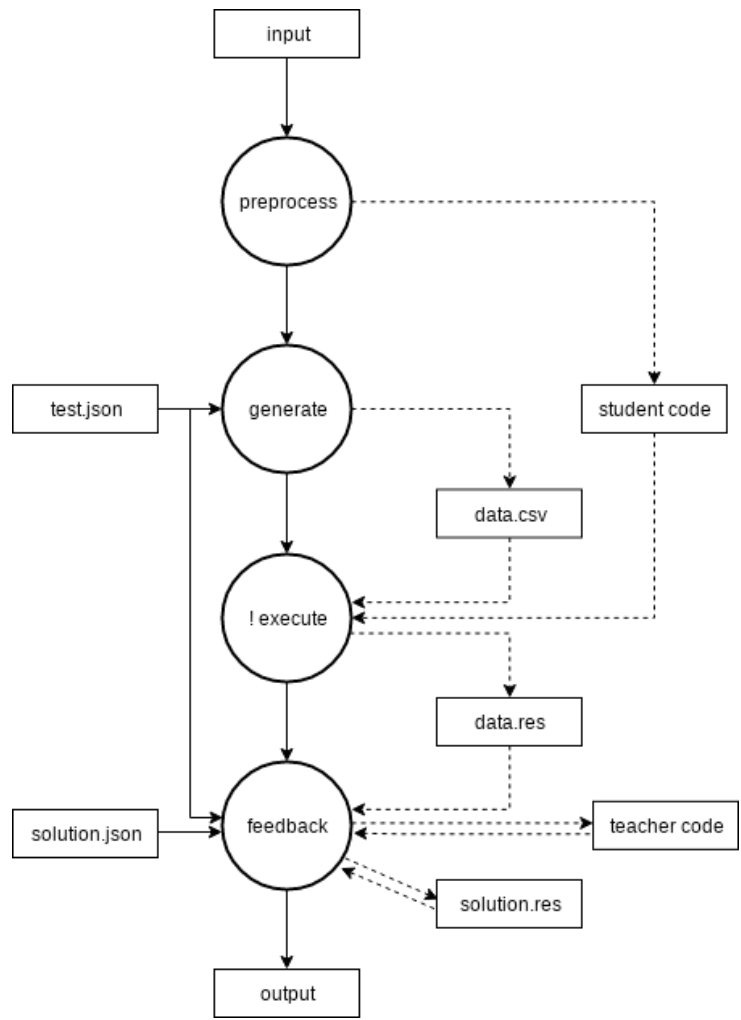

Figure 3. The structure of a unit testing-based task is composed of four main processes, namely the pre-processing, the tests suite generation, the code execution and the feedback generation.

Following this general structure for a unit testing-based task, it is possible to automatically generated exercises just providing some configuration information, described in the following section. Moreover, the only parts that are language-dependent are the execution of the student and the teacher code, all the rest being language-agnostic. To ease the implementation of unit testing-based tasks, the language-agnostic parts have been implemented as a independent tool written with the Go programming language, so that to be efficient. The language-dependent parts are implemented as libraries written in the target language for the exercise. For now, those libraries have only been written for the Python and Java programming languages.

## 3. ASSESSMENT STRUCTURE

To create a new exercise following the unit testing-based task template, an instructor has just to provide some basic configuration information structured as one JSON file, such as the one shown on Figure 4. The configuration consists in three distinct parts: (a) the specification, (b) the tests and (3) the solution. Except for the correct solution, all the other parts are language- agnostic and analysed either by the Go tool or by the language-dependent library. This task example asks the learner to write the body of a function sub that takes two parameters a and b, and that should return their subtraction, that is, a - b.

```json
{
    "spec": {
        "name": "sub",
        "args": {
            {
                "name": "a",
                "type": "int"
            },
            {
                "name": "b",
                "type": "int"
            }
        ],
        "return": "int"
    },
    "test": {
        "predefined": [
            {
                "data": "(10, 5)",
                "feedback": {
                    "10": "Have you subtracted the 2nd parameter?"
                }
            },
            {
                "data": "(7, 15)"
            },
            {
                "data": "(-1, 2)",
                "feedback": {
                    "*": "Have you considered negative parameters?"
                }
            },
            {
                "data": "(12, 0)"
            }
        ],
        "random": {
            "n": 10,
            "args": [
                "int(-20,20)",
                "int(-20,20)"
            ]
        }
    },
    "solution": {
        "f1": "return a - b"
    }
}
```

Figure 4. A unit testing-based task can be generated from a configuration file containing information about the specifications of the function to write, information about the predefined and random tests to be executed and finally one correct solution for the task.

The configuration file consists of three parts:

- The specification part (spec) is used to generate the code templates from which the student and teacher codes will be generated thanks to the input submission from the learner and the correct solution from the instructor. It contains all the information related to the signature of the function that the learner has to implement for the task.

- The tests part (test) contains predefined tests that have to be run and information and constraints used to generate random tests. It also contains information about customised feedback message that can be produced to help the learner if he/she fails the test.

- Finally, the solution part (solution) contains one possible solution for the task. It consists of chunks of code that are used to generate the teacher solution that is executed to get the correct answers for the test suite.

The POST /api/tasks route of the API server takes several parameters among which the type of the task to create can be specified (unit-testing for unit testing-based tasks) along with the configuration (such as described by Figure 4) and the programming language. The task generator then builds a Pythia task with all this information, using the language-agnostic code for the pre-process, generate and feedback components and the language-specific code for the execute component. Thanks to this feature, an instructor can generate a coding exercise without having to write any line of code. A user interface can be designed to help instructor design such exercise visually, wrapping the creation of the JSON configuration file and the call to the API server. An experiment conducted by the EDITx private company is currently underway, asking higher education professors in charge of introductory programming courses at the bachelor level to write unit testing-based exercises thanks to the proposed platform.

## 4. CONCLUSIONS

The tool presented in this paper is the result of further development of the version of [7], which now has the ability to automatically generate unit testing-based exercises that can be automatically graded. An instructor willing to design an exercise does not have to write any lines of code, except to provide one correct solution for the exercise. The presented tool combines advantages from competition, TDD and education graders so that to be used for education and learning purpose. It can also generate "intelligent" feedbacks to support learning, providing the learner with hints about his/her faults. The automatic generation of exercises process has been designed to be easy which should encourage instructors to create more exercises for their learners. It should also encourage easier sharing between educators.

Of course, the main strength of the Pythia platform being its high flexibility, future developments of the platform include the addition of new kinds of exercises, with the automatic grading and the automatic generation parts. Writing a task for the platform is not easy, but thinking about a generic kind of task, from which instances can be easily created, without having to write any line of code is even less easy but way more interesting. Some insights about how to include input-output tasks to the Pythia platform, with the automatic grading and generation parts have already been found. The next feature will be the addition of those kind of exercises, where the instructor only provide a statement along with a set of string inputs with the corresponding expected string output. For such exercises, the instructor will no longer have to provide any line of codes to design a new task, since he/she will not even have to provide any correct solution.

The platform is currently being used for several courses at the ECAM Brussels Engineering School and on the IT challenges platform of the EDITx private company. Informal evaluations from usage of previous versions of the platform already showed that the platform does bring useful help to learners. Future work includes a more rigorous evaluation of the platform and, in particular, should analyse the experiments in progress. Also, research has to be conducted to formally measure if the produced feedback information does indeed improve the learning performance of learners. It should also evaluate if the exercise creation process is easy and convenient enough for instructors.

## AUTHORS

**Dr Sébastien Combéfis** obtained his PhD in engineering in November 2013 from the Université catholique de Louvain (UCLouvain). He is currently working as a lecturer at the ECAM Brussels Engineering School, where his courses focus on computer science. He also obtained an advanced master in pedagogy in higher education in June 2014. Co-founder of the Belgian Olympiad in Informatics (be-OI) in 2010, he later introduced the Bebras contest in Belgium in 2012 and at the same time founded CSITEd. This non-profit organisation aims at promoting computer science in secondary schools.

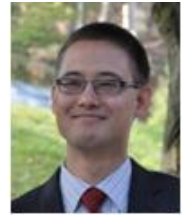

**Guillaume de Moffarts** is a master student in computer science at Université catholique de Louvain (UCLouvain). He is interested in computer science and electronics, and very curious about engineering and new technologies, such as 3D printing, artificial intelligence and the internet of things. He is also involved in the CSITEd non-profit organisation, taking part on several projects it organises. He was also recently the deputy leader of a Belgian delegation to the IBU Olympiad in Informatics 2019 that was held in Skopje, North Macedonia.

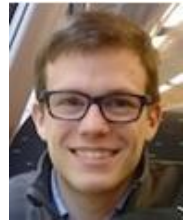